\documentclass[aps,prl,superscriptaddress,showpacs,twocolumn]{revtex4}
\usepackage{epsfig,units}
\begin{document}
\title{g-Factor Tuning and Manipulation of Spins by an Electric Current}
\author{Zbyslaw Wilamowski}
\affiliation{Institut f{\"u}r Halbleiter- und Festk{\"o}rperphysik, Johannes Kepler Universit{\"a}t, A-4040 Linz, Austria}
\affiliation{Institute of Physics, Polish Academy of Sciences, Al.\ Lotnikov 32/46, PL 02-668, Warsaw, Poland}
\author{Hans Malissa}
\affiliation{Institut f{\"u}r Halbleiter- und Festk{\"o}rperphysik, Johannes Kepler Universit{\"a}t, A-4040 Linz, Austria}
\author{Friedrich Sch{\"a}ffler}
\affiliation{Institut f{\"u}r Halbleiter- und Festk{\"o}rperphysik, Johannes Kepler Universit{\"a}t, A-4040 Linz, Austria}
\author{Wolfgang Jantsch}
\affiliation{Institut f{\"u}r Halbleiter- und Festk{\"o}rperphysik, Johannes Kepler Universit{\"a}t, A-4040 Linz, Austria}
\date{\today}
\begin{abstract}
We investigate the Zeeman splitting of two-dimensional electrons in an asymmetric silicon quantum well, by electron-spin-resonance (ESR) experiments. Applying a small dc current we observe a shift in the resonance field due to the additional current-induced Bychkov-Rashba (BR) type of spin-orbit (SO) field. This finding demonstrates SO coupling in the most straightforward way: in the presence of a transverse electric field the drift velocity of the carriers imposes an effective SO magnetic field. This effect allows selective tuning of the g-factor by an applied dc current. In addition, we show that an ac current may be used to induce spin resonance very efficiently.
\end{abstract}
\maketitle
Dirac's theory of the free electron implies already that the Zeeman splitting in a magnetic field is modified for particles with finite momentum due to the relativistic increase of the electron mass. The free electron g-factor extrapolates to a value close to 2 only in the limit of small velocity. In atoms and solids, spin-orbit (SO) interaction may be ascribed to the magnetic field originating from the motion of an electron in the electric field of the other charges \cite{ref1}.

In solids, the lowest order -- bilinear -- SO term for the energy of an electron contains the vector product of the electron velocity (or equivalently its momentum, $\hbar\vec{k}$) and the spin, $\sigma$. Such a linear term can only exist for lower than mirror symmetry, induced e.g., by the existence of an electric field, $\vec{E}_{0}$, as discussed by Bychkov and Rashba \cite{ref2,ref3}. This type of SO coupling is described by a term $\hbar\omega_{BR}=\alpha_{BR}\left(\vec{k}\times\hat{n}\right)\cdot\stackrel{\leftrightarrow}{\sigma}$, which causes spin splitting already without an externally applied magnetic field. Here $\alpha_{BR}$ is the Bychkov-Rashba coefficient which depends on the strength of SO interaction and the asymmetry of the system, $\stackrel{\leftrightarrow}{\sigma}$ stands for the Pauli matrix describing the spin state, and $\hat{n}$ is a unit vector pointing in the direction in which the symmetry is broken (parallel to $\vec{E}_{0}$).

\begin{figure}[bth]
\epsfig{width=.7\columnwidth,file=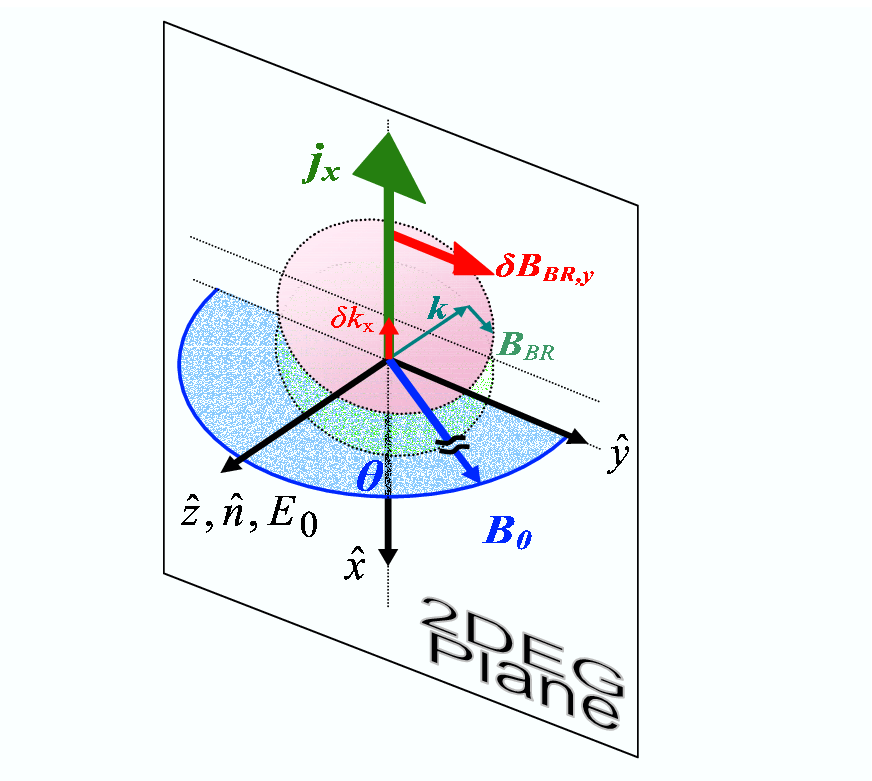}
\caption{\label{fig1} A current $j_{x}$ passes through a 2D electron gas ($x$-$y$ plane). The Fermi circle (momentum distribution, green for $j_{x}=0$), shifts by an amount $\delta k_{x}$ (purple, $j_{x}\ne 0$). Within this approximation, each electron experiences a Bychkov-Rashba field $\delta B_{BR,y}$ in addition to that resulating from its momentum $\hbar k$. The static field $B_{0}$ (drawn not to scale) is applied in the $y$-$z$ plane to enable ESR measurements.}
\end{figure}

The SO-induced spin splitting can be described also in terms of an effective magnetic ``Bychkov-Rashba'' (BR) field, $\vec{B}_{BR}=\alpha_{BR}\left(\vec{k}\times\hat{n}\right)/g\mu_{B}$, seen by each electron. The BR field is perpendicular to both the momentum of the electron, $\hbar\vec{k}$, and the direction $\hat{n}$, and thus is oriented in-plane and perpendicular to the electron momentum. In Fig.~\ref{fig1}, $\vec{B}_{BR}$ is indicated by an arrow for a particular electron moving with a momentum vector $\vec{k}$ in a two-dimensional electron gas (2DEG).

For a 2DEG in a semiconductor quantum well, a perpendicular built-in electric field, $\vec{E}_{0}$, may be imposed, e.g., by one-sided modulation doping -- a method by which free carriers are introduced to the quantum well by accommodating donors at a distance of some \unit[10]{nm} from the quantum well in the barrier material. This arrangement suppresses both ionized impurity scattering and carrier freeze-out at low temperatures.

Various effects of the BR field have been demonstrated already. In a 2DEG, the BR field causes anisotropy of both the ESR line width and the line position \cite{ref4,ref5}. The BR field has been shown also to cause additional longitudinal spin relaxation of the Dyakonov-Perel type \cite{ref5,ref6}. In equilibrium, the mean value of the thermal velocity vanishes and thus these effects are only of second order: they are proportional to the square of the BR field at the Fermi level, $B_{BR}^{2}$. Here we consider the effect of a macroscopic current of density $j_{x}$ within a 2DEG. The non-vanishing mean carrier velocity leads to a first-order BR field, $\delta B_{BR,y}$, which causes additional spin splitting and thus a shift in the electron spin resonance (ESR).

Other effects of an electric current on the spin properties have been considered before. In their seminal paper, Datta and Das proposed to make use of spin precession around the BR field to build a structure with controllable resistance \cite{ref7}. The BR field, due to spin-dependent relaxation, causes also spin polarization \cite{ref8,ref9,ref10}. The so-called spin galvanic effects belong to the same class of phenomena \cite{ref11}. In the spin-galvanic effect a macroscopic current is caused by different relaxation rates of photo-induced carriers with opposite spin. In the inverse spin galvanic effect the asymmetry of spin relaxation rates for electrons in the two spin sub-bands split by SO coupling leads to a spin polarization induced by an applied electric current.

We believe, however, that the effect presented in this paper demonstrates the occurrence of the BR field in the most direct way and it also has practical implications for the realization of spin-based electronic devices. At the end of this paper we show that a high frequency (\emph{hf}) current, which may be induced by the microwave fields in an ESR experiment, leads to an \emph{hf} BR field. The latter may exceed the applied microwave magnetic field substantially and thus it can be utilized for a most efficient spin manipulation.

We investigate the ESR of the conduction electrons in an MBE-grown Si quantum well defined by Si$_{0.75}$Ge$_{0.25}$ barriers. The layer structure and the basic ESR properties have been described elsewhere \cite{ref5,ref12,refnew}. Here we added electric contacts to the 2DEG. The sample was then glued to a quartz holder containing also wires connected to the 2DEG and inserted into a TE102 rectangular microwave cavity equipped with an intracavity cryostat, which allows cooling to \unit[2.5]{K}. ESR measurements were performed with a standard X-band Bruker ElexSys E500 system.

\begin{figure}[bth]
\epsfig{width=\columnwidth,file=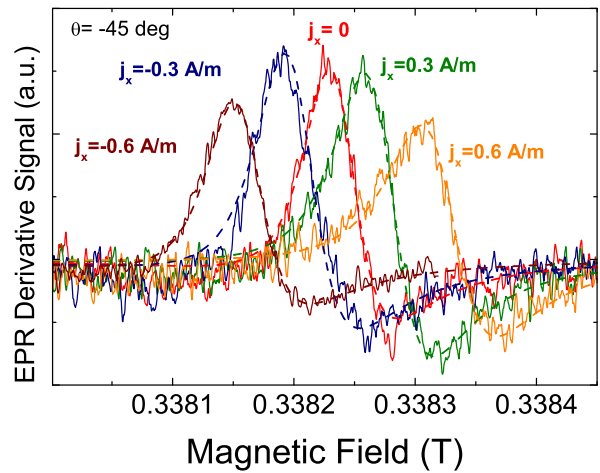}
\caption{\label{fig2} ESR spectra of a 2DEG in a Si quantum well for various values of an electric current density passing a \unit[3]{mm} wide sample. Measurements were performed with $B_{0}$ tilted by $\theta=-45^{\circ}$ from the direction perpendiclar to the sample plane at a microwave frequency of \unit[9.4421]{GHz}.}
\end{figure}

Spectra are given in Fig.~\ref{fig2}. Due to the use of field modulation and lock-in detection (standard in ESR instruments) we obtain the first derivative of the microwave absorption with respect to $B_{0}$. Here the static magnetic field $\vec{B}_{0}$ was tilted by $\theta=45^{\circ}$ with respect to the sample surface normal, $\hat{n}$. The line shape is asymmetric, which will be discussed below in the context of the \emph{hf} current induced ESR. Spectra are given for different dc currents applied during the measurement. It is clearly seen that (i) a current shifts the resonance, (ii) the shift occurs in the opposite sense when the current direction is inverted and (iii) the signal broadens with increasing current.

\begin{figure}[bth]
\epsfig{width=\columnwidth,file=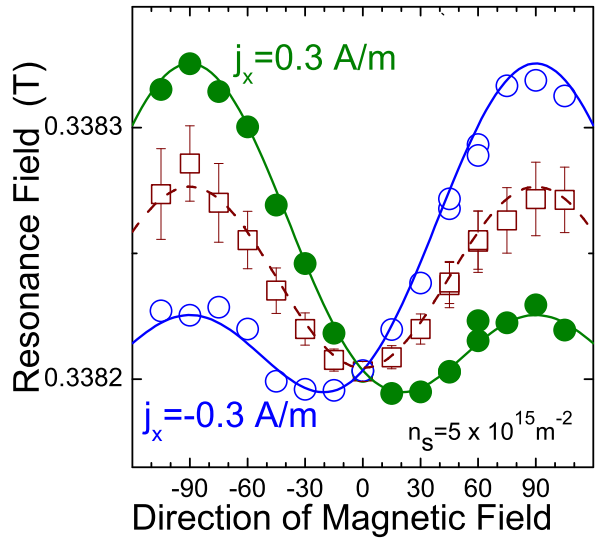}
\caption{\label{fig3} Angular dependence of the ESR field for a current $J=0$ (squares) and $\unit[\pm 1]{mA}$ (open and full circles, respectively). The electron concentration is $n_{S}=\unit[5\cdot 10^{15}]{m^{-2}}$ and the sample width \unit[3]{mm}. Error bars correspond to 20\% of the resonance line width.}
\end{figure}

In thermal equilibrium ($j_{x}=0$), the anisotropy of the ESR position can be fully described by treating $\vec{B}_{BR}$ like a real field in the range considered here and by adding it to the external field \cite{ref4,ref5}. In spite of the isotropic distribution of the Fermi momenta and the resulting isotropic distribution of the BR fields, their superposition with the external field $B_{0}$ results in the angular dependence shown by the open squares in Fig.~\ref{fig3}. This anisotropy allows the evaluation of the mean value \cite{ref13} of $\left<B_{BR}^{2}\right>$ at the Fermi circle and of $\alpha_{BR}$. The dashed line corresponds to a fit using $B_{BR}=\unit[10]{mT}$. For this sample with an electron concentration of $n_{S}=\unit[5\cdot 10^{15}]{m^{-2}}$ this value yields a BR coefficient of $\alpha_{BR}=\unit[0.85\cdot 10^{-12}]{eVcm}=\unit[1.4\cdot 10^{-33}]{Jm}$, which compares well to earlier published values \cite{ref4}.

Now we pass a current through the sample. A current causes an antisymmetric shift of the ESR position as seen in Fig.~\ref{fig3}. For in-plane field ($\theta=90^{\circ}$) the current-induced shift is maximum as $\delta B_{BR}$ is oriented along the $y$-axis, parallel to $B_{0}$ (s.\ Fig.~\ref{fig1}) and thus the external field required for resonance is reduced by $\delta\vec{B}_{BR,y}$. For $\theta=-90^{\circ}$, $\delta\vec{B}_{BR,y}$ is antiparallel to the applied field and therefore the resonance field is increased by the same amount.  For the data presented the maximum shift is \unit[50]{$\mu$T} and this shift directly corresponds to $\delta B_{BR,y}$. For $\theta=0^{\circ}$, $\delta\vec{B}_{BR,y}$ is perpendicular to the applied field and its effect on the resonance field is negligible. Altogether, the observed geometrical dependencies reflect the vector product of electron velocity and the built-in electric field, which characterizes SO coupling.

For moderate electric fields, the current density, $j_{x}$, can be described in terms of a drift shift of the Fermi circle by: $\delta k_{x}=-m^{*}j_{x}/e\hbar n_{S}$ (see Fig.~\ref{fig1}). Consequently, since $B_{BR}$ increases with increasing k-vector, each electron experiences an additional BR field $\delta B_{BR,y}$ due to $\delta k_{x}$ (see Fig.~\ref{fig1}). This additional field is proportional to the shift of the Fermi circle, yielding $\delta B_{BR,y}=\beta_{BR}v_{d}$, where $\beta_{BR}=\alpha_{BR}m^{*}/g\mu_{B}\hbar$ is a material parameter and $v_{d}$ stands for the drift velocity.

\begin{figure}[bth]
\epsfig{width=\columnwidth,file=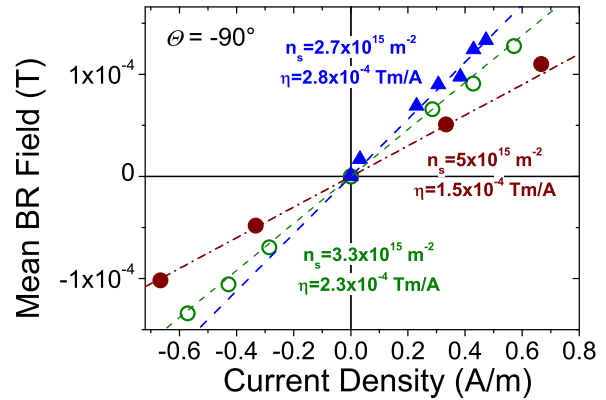}
\caption{\label{fig4} Dependence of the resonance shift for in-plane orientation of $\vec{B}_{0}$ on the dc electric current density for three samples of different sheet carrier concentration, $n_{S}$.}
\end{figure}

The current-induced resonance shift is thus expected to change linearly with current, where the slope $\eta=\delta B_{BR,y}/j_{x}=\beta_{BR}/en_{S}$ is proportional to $\alpha_{BR}$ and inversely proportional to $n_{S}$. Our experiments confirm this as shown in Fig.~\ref{fig4}. The dependence of the resonance shift (for in-plane orientation) on the current density is shown for 3 samples with different electron concentrations. The constant slope, $\eta$, is larger for smaller $n_{S}$. The experimental value of $\eta$ allows for an independent evaluation of the BR parameter, $\alpha_{BR}$. Within the experimental error of about 20\%, all values for $\alpha_{BR}$, as obtained from the ratio $\eta$, are equal to those obtained from the anisotropy of the resonance field in the absence of an electric current and independent of temperature.

From the observed effect of a dc current we may also infer that high frequency effective fields can be generated by an \emph{hf} current. The latter is limited in frequency only by the momentum scattering rate and therefore effective microwave magnetic fields can be generated this way. High frequency fields are of particular interest as they can be used to excite ESR. Using microwave pulses of specific duration, the magnetization can be turned by any Rabi angle, as has been demonstrated in spin-echo experiments \cite{ref14}.

In a classical ESR experiment, we are looking for magnetic dipole transitions. Therefore the sample is placed in the node of the electric field within a microwave cavity in order to minimize losses due to conductance effects. Nevertheless, microwave currents within the high mobility 2DEG are evident in our experiment from the appearance of cyclotron resonance \cite{ref5} which may cause electric currents and thus additional BR fields.

We find evidence for the \emph{hf} BR field from an analysis of the ESR amplitude and line shape. The latter (see Fig.~\ref{fig2}) shows some asymmetric, dispersive component, similar to the Dysonian line shape discussed for 3D metals \cite{ref15a,ref15b}. Modification of Dyson's model for 2D samples yields, however, only a pure absorption signal. Here the phase shift of the \emph{hf} BR field with respect to the microwave electric field must be taken into account to explain the line shapes observed. Our quantitative modeling explains the relative magnitude of dispersive and absorption signals as a function of microwave power and geometry \cite{ref16}. We conclude that the appearance of a dispersive signal component in a 2DEG is a strong indication for a current-induced BR field.

Experimentally we investigated also the influence of the orientation of the 2DEG relative to $\vec{B}_{0}$ and the microwave magnetic field, $\vec{B}_{1}$, which is oriented in the $x$-direction in Fig.~\ref{fig1}. Rotating the sample within the cavity about the $x$ and $z$ directions, we find qualitative agreement with the model. The highest signal is obtained if the 2DEG is perpendicular to $\vec{B}_{1}$. In that case, $\vec{B}_{1}$ very efficiently induces eddy currents within the 2DEG, which in turn cause an additional microwave magnetic BR field that is much stronger than the original microwave magnetic field. The expected gain in ESR excitation, and the resulting Rabi frequency is proportional the electron mobility in the 2DEG and for state of the art mobilities in Si quantum wells we estimate gain values of $10^{3}$ and more.

The demonstrated effect of a current-induced spin resonance is somehow similar to the known effect of the electric dipole spin resonance \cite{ref3,ref17}. Both effects originate from a time modulation of the SO field. Electric dipole transitions originate, however, from the modulation of the electric field while the current-induced effect comes from the modulation of the carrier velocity.

The presented experimental data demonstrate the occurrence of a current-induced spin orbit field. A dc current allows tuning of the ESR frequency (and thus the g-factor), while a high frequency current occurs to be a very effective tool for spin excitation, or generally, for spin manipulation. This method of spin tuning and manipulation can be applied locally, e.g., to a nano-wire without heating of the rest of a sample in contrast to methods employing a resonator.

Both the Rabi frequency and the spin relaxation rate increase with increasing SO coupling. SO coupling in III-V compounds is by up to three orders of magnitude stronger than in Si. Here the Rabi frequency scales linearly with the SO interaction and the line width with the square of it. Therefore, materials such as Si are much better suited if a large shift-to-line width-ratio of the ESR is needed. On the other hand, we expect a higher efficiency for the current-induced spin manipulation for III-V compounds.

The current-induced shift of the spin resonance described in this paper is probably the most direct and conceptually simplest effect of SO interaction in solids. Moreover, the ratio of the g-shift and current density is ruled by the BR parameter and the carrier density only, but it is independent of temperature, electron mobility or details of spin relaxation.
\begin{acknowledgments}
This work was supported by the \emph{Fonds zur F{\"o}rderung der Wissenschaftlichen Forschung} (Project P16631-N08), and the {\"O}AD, both Vienna, Austria, and in Poland by KBN.
\end{acknowledgments}


\begin{thebibliography}{18}
\bibitem[{ref()}]{ref13}
\bibinfo{note}{The characteristic time in measuring ESR is the inverse
  linewidth. Here it is of the order of \unit[1]{$\mu$s}, i.e., very long as
  compared to the momentum relaxation time of about \unit[$10^{-11}$]{s}. As a
  result, the ESR data yield the mean BR field of the ensemble rather than the
  BR field seen by individual electrons (see also Ref.\ \cite{ref6}).}
\end{thebibliography}
\end{document}